\documentclass[Physsubmission, Phys]{SciPost}

\hypersetup{
    colorlinks,
    linkcolor={red!50!black},
    citecolor={blue!50!black},
    urlcolor={blue!80!black}
}

\usepackage[bitstream-charter]{mathdesign}
\DeclareSymbolFont{usualmathcal}{OMS}{cmsy}{m}{n}
\DeclareSymbolFontAlphabet{\mathcal}{usualmathcal}

\begin{document}

\begin{center}{\Large \textbf{
Recent CMS results on soft QCD physics\\
}}\end{center}

\begin{center}
Rajat Gupta\textsuperscript{$\star$} (on behalf of the CMS Collaboration)
\end{center}

\begin{center}
Panjab University, Chandigarh
\\
* rajat.gupta@cern.ch
\end{center}

\begin{center}
\today
\end{center}


\definecolor{palegray}{gray}{0.95}
\begin{center}
\colorbox{palegray}{
  \begin{tabular}{rr}
  \begin{minipage}{0.1\textwidth}
    \includegraphics[width=30mm]{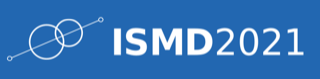}
  \end{minipage}
  &
  \begin{minipage}{0.75\textwidth}
    \begin{center}
    {\it 50th International Symposium on Multiparticle Dynamics}\\ {\it (ISMD2021)}\\
    {\it 12-16 July 2021} \\
    \doi{10.21468/SciPostPhysProc.?}\\
    \end{center}
  \end{minipage}
\end{tabular}
}
\end{center}

\section*{Abstract}
{\bf
Soft quantum chromodynamics (QCD) measurements play an important role in fundamental QCD studies as well as in tuning of corresponding Monte Carlo generator models for a good description of experimental data. Recent results of soft QCD measurements with the CMS experiment, such as minimum bias/underlying event physics, double parton scattering and forward jet production production are presented. 
}

\vspace{10pt}
\noindent\rule{\textwidth}{1pt}
\tableofcontents\thispagestyle{fancy}
\noindent\rule{\textwidth}{1pt}
\vspace{10pt}

\section{Introduction}
\label{sec:intro}

The dominant component of proton-proton (pp) collision comes from the processes involving soft and semi-hard quantum chromodynamics (QCD). Soft interactions cannot be described in terms of perturbative QCD, so it requires phenomenological models for their description. In this document, latest CMS~\cite{CMS:2008xjf} and TOTEM~\cite{TOTEM:2008lue} measurements on soft QCD physics and their comparisons with predictions of various theoretical models are presented.

\section{DPS study using inclusive four jets process}

Because of the complex structure of nucleons, many parton-parton interactions can occur in a single pp collision. Double parton scattering (DPS) refers to events in which two hard parton-parton interactions occur in single pp collision. A study of inclusive production of four-jets in pp collisions at a center-of-mass energy ($\sqrt{s}$) of 13 TeV is presented~\cite{CMS:2021ijt}. 

The study is performed using the data, which correspond to an integrated luminosity of 0.042 $pb^{-1}$, collected using the CMS detector at the CERN LHC in 2016, during a special data taking period with a low probability for several pp interactions occurring within the same or neighbouring bunch crossings (hence referred to as `pileup'). Differential cross section is measured as functions of jet $p_{T}$, jet pseudorapidity ($\eta$), and several other observables that describe the angular correlations between the jets. 

Two phase space regions are choosen, each determined by choices on jet $p_T$.  Region I is defined by four leading jets which are required to lie within  $|\eta| <$ 4.7, with $p_T$ thresholds of 35, 30, 25, and 20 GeV.  Asymmetric selection is chosen over symmetric ones, since the symmetric threshold tends to reduce the DPS contribution with respect to the single parton scattering (SPS) fraction, according to higher-order calculations or calculations performed in the $k_T$-factorization framework. In region II, the $\Delta S$ distribution is obtained with $p_T$ thresholds of 50, 30, 30, and 30 GeV. Since there are no jet triggers with $p_T$ thresholds below 30 GeV, this selection is required to obtain the cross sections $\sigma_A$ and $\sigma_B$ used in the extraction of $\sigma_{eff}$. The measured distributions are corrected for detector effects with the TUnfold program and compared with \textsc{pythia}~8, \textsc{herwig}7, \textsc{herwig}++ and multijet models.


Different parton shower implementations have less impact on the $\Delta S$ distribution. The DPS tune CDPSTP8S1-4j describes the shape within uncertainty, whereas all other models predictions underestimate the data at low $Delta S$, indicating a possible need for more DPS contribution. 



The contribution of the DPS is extracted using a template fitted to the data, SPS distributions derived from Monte Carlo event generators,  as well as a DPS distribution obtained from inclusive single-jet events in data. The results of $\sigma_{\rm eff}$ derived with the models based on the current CP5 and CH3 tunes and where the hard MPI has been eliminated are shown in figure~\ref{fig:2b}.

All results, except for the values obtained with the NLO $2 \to 2$ models, agree with the measurement performed by the ATLAS collaboration at a $\sqrt{s}$ of 7~TeV, where a $\sigma_{\rm eff}$ equal to $14.9^{+1.2}_{-1.0}\mathrm{(stat)}^{+5.1}_{-3.8}\mathrm{(syst)}~\mathrm{mb}$ was found, while none agree with the value of $21.3^{+1.2}_{-1.6}~\mathrm{mb}$ from the CMS measurement at a $\sqrt{s}$ of 7~TeV, which is more in line with the results obtained with some of the models based on older underlying event (UE) tunes.  

\section{DPS study using inclusive Z+jets process}

Another DPS measurement is performed to explore various observables which are sensitive to the DPS using Z+jets process with the CMS detector at $\sqrt{s}$ = 13 TeV, where the Z boson decays into two oppositely charged muons~\cite{CMS:2021wfx}. In order to be consistent with previous DPS measurements, jets are considered to have a $p_T$ threshold of 20 GeV. 

The measured production cross section in the fiducial region is found to be 58.5 $\pm$ 0.3 (stat) $\pm$ 7.0 (syst) $\pm$ 1.2 (theo) $\pm$ 4.0 (lumi) pb for Z + $\geq$ 1 jet events and 44.8 $\pm$ 0.4 (stat) $\pm$ 3.7 (syst) $\pm$ 0.5 (theo) $\pm$ 1.1 (lumi) pb for Z $+ \geq$ 2 jets events. The cross section is well described by various simulation within uncertainties, except MG5\_aMC + \textsc{pythia}~8 (with DPS specific CDPSTP8S1-WJ tune), which overestimates the measurement by 10\%.

\begin{figure}[htbp]
\begin{minipage}[t]{.48\textwidth}
\centering
\includegraphics[width=1\textwidth]{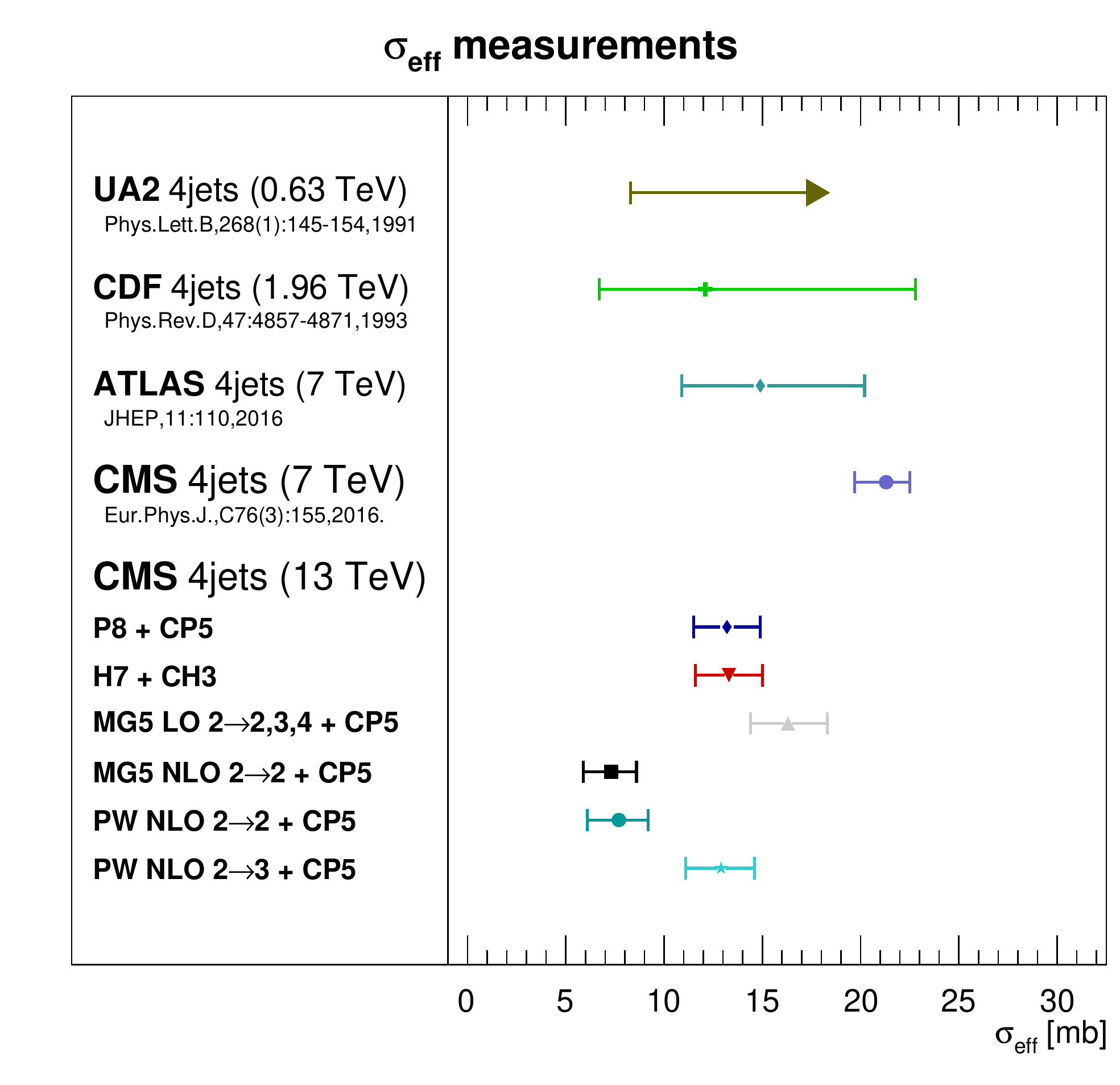}
\caption {{Comparison of the values for $\sigma_{\rm eff}$ extracted from data. The results from four-jet measurements performed at lower $\sqrt{s}$ are shown alongside the newly extracted values. Reproduced from Reference \cite{CMS:2021ijt} (CC BY 4.0).}} \label{fig:2b}
\end{minipage}
\hfill
\begin{minipage}[t]{.48\textwidth}
\centering
\includegraphics[width=1\textwidth]{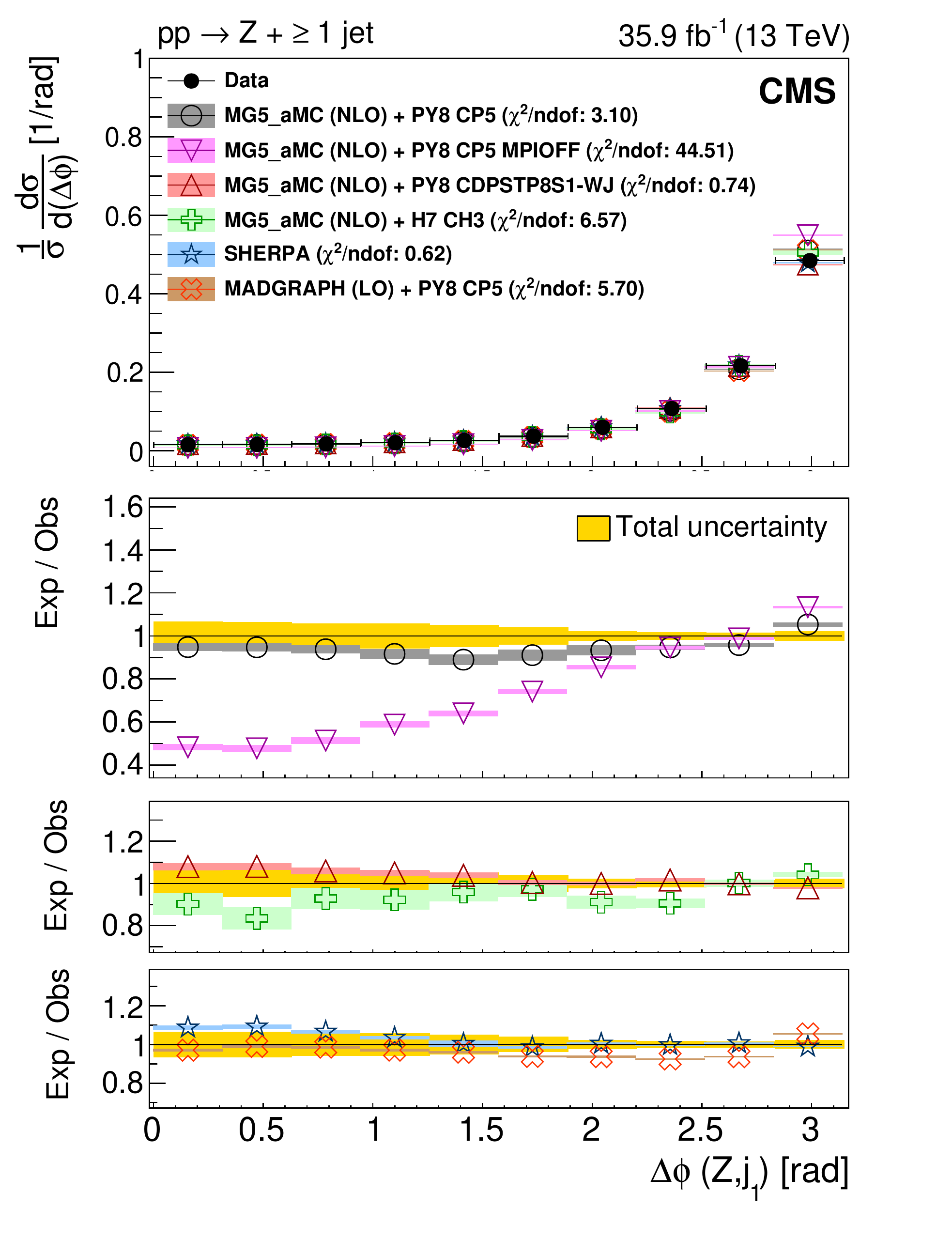}
\caption {{Differential cross sections as a functiona of $\Delta\phi$ between the Z boson and the leading jet for Z + $\geq$ 1 jet events. In the bottom panels, the total uncertainty for data is indicated by the solid yellow band centred at 1. Reproduced from Reference \cite{CMS:2021wfx} (CC BY 4.0).}} \label{fig:3}
\end{minipage}
\end{figure}

The differential cross section as a function of $\Delta\phi$ between the Z boson and the leading jet is shown in figure~\ref{fig:3}. The prediction of MG5\_aMC + \textsc{pythia}~8 without MPI underestimates the measurement by 50\% at lower $\Delta\phi$, which indicates the sensitivity of this distribution to the presence of MPI. Different MC predictions describes the the differential cross section distribution within uncertainties, except for the MG5\_aMC + \textsc{pythia}~8 with DPS-specific tune CDPSTP8S1-WJ, which deviates up to 10--20\%, but describes the shape of the observable (not shown in Figure). The results presented will be an important input for the development of DPS-specific tunes and global tunes combined with other soft QCD measurements in pp collisions on the TeV scale.

\section{Hard color singlet exchange in dijet events}

\begin{figure}[htbp]
\begin{minipage}[t]{.48\textwidth}
\centering
\includegraphics[width=1\textwidth]{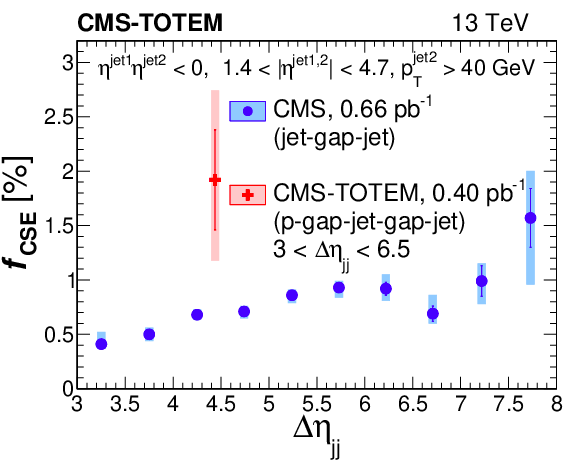}
\end{minipage}
\hfill
\begin{minipage}[t]{.48\textwidth}
\centering
\includegraphics[width=1\textwidth]{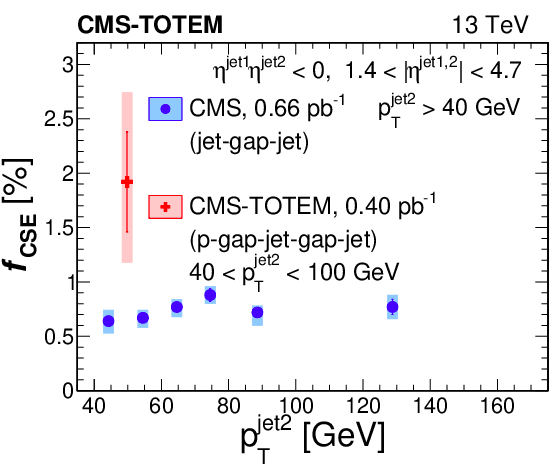}
\end{minipage}
\caption {{Gap fraction, $f_{\rm CSE}$, measured as a function of $\Delta\eta_{\rm jj}$ and $p_{\rm T}^{\rm jet2}$ in inclusive dijet events and in dijet events with a leading proton. Reproduced from Reference \cite{6} (CC BY 4.0).}}\label{fig:4}
\end{figure}

One of the processes sensitive to Balitsky-Fadin-Kuraev-Lipatov (BFKL) dynamics~\cite{bfkl} is the production of two jets separated by a large rapidity interval devoid of particle activity, known as Mueller-Tang jets~\cite{5} or jet-gap-jet events. The CMS detector with its unprecedented $\sqrt{s}$ and large detector coverage in rapidity provides an ideal tool for testing BFKL dynamics
and understanding the role of diffraction at large momentum transfers in strong nuclear interactions. The study~\cite{6} is performed with the low instantaneous luminosity data collected in pp collisions at $\sqrt{s} =$ 13 TeV by the CMS and TOTEM experiments. The fraction of jet-gap-jet events to events where the two jets have similar kinematics, $f_{\rm CSE}$, is measured as a function of the $\eta$ difference between the leading two jets, $\Delta\eta_{\rm jj}$, and the transverse momentum ($p_{\rm T}$) of the sub-leading jet ($p_{\rm T}^{\rm jet2}$) (Figure~\ref{fig:4}). An increase with $\Delta\eta_{\rm jj}$ and a weak dependency on $p_{\rm T}^{\rm jet2}$ are observed. The
present analysis sets a constraint on the theoretical treatment of rapidity gap survival probability.

\section{Central exlusive $\pi^+\pi^-$ production in pp collisions}

Central exclusive and semiexclusive production of $\pi^+\pi^-$ pairs is measured with the CMS detector in pp collisions at the LHC at $\sqrt{s}$ of 5.02 and 13 TeV~\cite{7}. Exclusive events are selected by vetoing additional energy deposits in the calorimeters and by requiring two oppositely charged pions identified via their mean energy loss in the tracker detectors. 

\begin{figure}[htbp]
\begin{minipage}[t]{.48\textwidth}
\centering
\includegraphics[width=1\textwidth]{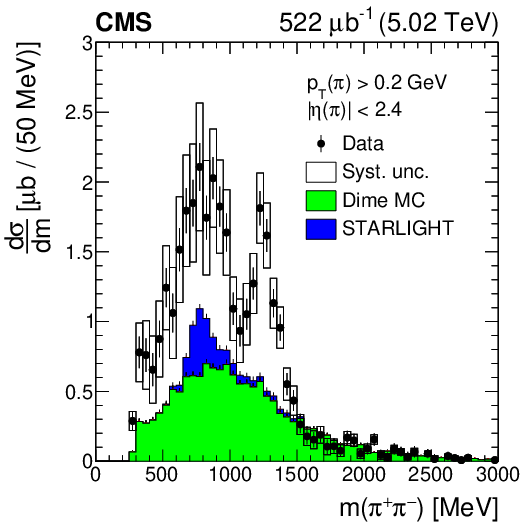}
\end{minipage}
\hfill
\begin{minipage}[t]{.48\textwidth}
\centering
\includegraphics[width=1\textwidth]{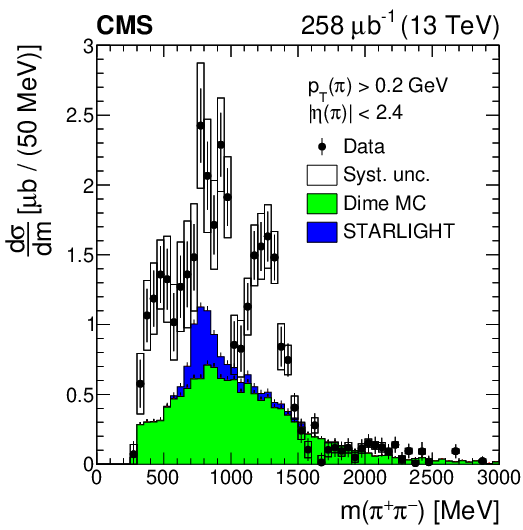}
\end{minipage}
\caption {{Differential cross sections as functions of mass compared with generator-level simulations for the 5.02 (left) and 13 TeV (right) data sets. Reproduced from Reference \cite{7} (CC BY 4.0
).}}\label{fig:5}
\end{figure}

These events are used together with correction factors to obtain invariant mass, $p_T$, and rapidity distributions of the $\pi^+\pi^-$ system. The total and differential cross sections of exclusive and semi-exclusive central $\pi^+\pi^-$ production are measured as functions of invariant mass, $p_T$, and rapidity of the $\pi^+\pi^-$ system in the fiducial region defined as $p_T$ ($\pi$) $>$ 0.2 GeV and pseudorapidity $|\eta(\pi)|$ $<$ 2.4. 

The observed mass spectrum (shown in Figure~\ref{fig:5} exhibits resonant structures, which can be fitted with a simple model containing four interfering Breit-Wigner functions, corresponding to the $f_0$ (500), $\rho^0$(770), $f_0$ (980), and $f_2$ (1270) resonances, and a continuum contribution modeled by the DIME MC. The exclusive production cross sections are extracted from this fit. The measured total exclusive $\pi^+\pi^-$ production cross section is 32.6 $\pm$ 0.7 (stat) $\pm$ 6.0 (syst) $\pm$ 0.8 (lumi) and 33.7 $\pm$ 1.0 (stat) $\pm$ 6.2 (syst) $\pm$ 0.8 (lumi) mb for 5.02 and 13 TeV, respectively. The obtained cross sections of $\rho^0$ (770) production are higher than the STARLIGHT model prediction, which can be explained by the presence of semiexclusive production which is not modeled by the STARLIGHT generator.

\section{Summary}

An outline of various soft QCD and diffractive measurements have been presented. An accurate modelling of soft QCD processes is vital for various complex analysis such as precision mass measurement of top quark. Although existing measurements already challenge different soft QCD models, there are still many possibilities to direct theory with new measurements.

\nolinenumbers

\end{document}